\documentclass[12pt,preprint]{aastex}

\shorttitle{Swift X-ray Study of HESS J0632+057}
\shortauthors{Falcone et al.}

\begin{document}


\title{Probing the Nature of the unidentified TeV Gamma-ray source HESS J0632+057 with {\it{Swift}}}


\author{
A.~D.~Falcone\altaffilmark{1,2},
J.~Grube\altaffilmark{3},
J.~Hinton\altaffilmark{4},
J.~Holder\altaffilmark{5},
G.~Maier\altaffilmark{6},
R.~Mukherjee\altaffilmark{7},
J.~Skilton\altaffilmark{4},
M.~Stroh\altaffilmark{1}
}

\altaffiltext{1}{Department of Astronomy \& Astrophysics, Pennsylvania State University, University Park, PA 16802, USA}
\altaffiltext{2}{corresponding author email: afalcone@astro.psu.edu}
\altaffiltext{3}{School of Physics, University College Dublin, BelÞeld, Dublin 4, Ireland}
\altaffiltext{4}{School of Physics and Astronomy, University of Leeds, Leeds LS2 9JT, UK}
\altaffiltext{5}{Dept. of Physics and Astronomy and the Bartol Research Institute, University of Delaware, Newark, DE 19716, USA}
\altaffiltext{6}{Physics Department, McGill University, Montreal, QC H3A 2T8, Canada}
\altaffiltext{7}{Department of Physics and Astronomy, Barnard College, Columbia University, NY 10027, USA}


\begin{abstract}
New generation TeV gamma-ray telescopes have discovered many new sources, including several enigmatic unidentified TeV objects. HESS J0632+057 is a particularly interesting unidentified TeV source since: it is a point source, it has a possible hard-spectrum X-ray counterpart and a positionally consistent Be star, it has evidence of long-term VHE flux variability, and it is postulated to be a newly detected TeV/X-ray binary. We have obtained Swift X-ray telescope observations of this source from MJD 54857 to 54965, in an attempt to ascertain its nature and to investigate the hypothesis that it's a previously unknown X-ray/TeV binary. Variability and spectral properties similar to those of the other 3 known X-ray/TeV binaries have been observed, with measured flux increases by factors of $\sim$3. X-ray variability is present on multiple timescales including days to months; however, no clear signature of periodicity is present on the timescales probed by these data. If binary modulation is present and dominating the measured variability, then the period of the orbit is likely to be $\ge$ 54 days (half of this campaign), or it has a shorter period with a variable degree of flux modulation on successive high states. If the two high states measured to date are due to binary modulation, then the favored period is approximately 35-40 days. More observations are required to determine if this object is truly a binary system and to determine the extent that the measured variability is due to inter-orbit flaring effects or periodic binary modulation.
 
\end{abstract}





\section{Introduction}
During the past 5 years, TeV gamma ray astronomy has experienced rapid growth due to the commissioning of a new generation of sensitive atmospheric Cherenkov telescopes. These telescopes have led to an expansion of the TeV catalog from $\sim$10 sources to nearly 100 sources in just a few years \citep{horan09,weekes09,buckley08,aharonian08}. Many of these sources have come from the survey of the Galactic plane by the HESS telescopes in Namibia. One of the most exciting results of this survey is a large number of unidentified TeV sources, which offer a new region of discovery space at very high energies. HESS J0632+057, located within the Monoceros Loop region, is one of these unidentified sources.

HESS J0632+057 was first detected by HESS using 13.5 hours of data collected between March 2004 and March 2006 \citep{aharonian07}. It has a peak significance in the field of view of 7.1$\sigma$, and the post-trials significance is 5.3$\sigma$, after accounting for the trials associated with searching the entire field of view. The spatial distribution is consistent with a point source (RMS size $<$ 2 arcmin at 95\% confidence) near the edge of the Rosette Nebula, located at RA 06h 32' 58.3'', Dec. +05$^{\circ}$ 48' 20'' ($\pm$ 28'' stat., 20'' sys.). The source position is located on the edge of the 99\% contour of the EGRET unidentified object 3EG J0634+0521, but no object at the position of HESS J0632+057 is listed in the the Fermi bright source list \citep{abdo2009}. MWC 148, a Be star, is at the centroid of the HESS position. \citet{aharonian07} suggested that the TeV gamma ray emission could be associated with either: a) a shock front driven by the stellar wind of MWC 148, b) shock accelerated cosmic rays interacting with a molecular cloud at the edge of the Monoceros Loop, c) an association with an unidentified ROSAT source 1RXS J063258.3+054857, d) an association with EGRET unidentified gamma ray source 3EG J0634+0521, and/or e) a previously unknown X-ray binary source associated with MWC 148. 

On 2007 September 17, XMM-Newton observed the region surrounding HESS J0632+057, obtaining 26 ksec of good data \citep{hinton09}. This resulted in the detection of several sources, the brightest being a point source, XMMU J063259.3+054801, positionally coincident with MWC 148 and HESS J0632+057 and was within the 99\% error circle of 1RXS J063258.3+054857. During the relatively short observation period, the object exhibited significant variability in the form of a gradual decline from $\sim$130 c/ksec to $\sim$90 c/ksec. During this observation, it had a mean deabsorbed 1-10 keV flux of (5.3 $\pm$ 0.4) $\times 10^{-13}$ erg~cm$^{-2}$ s$^{-1}$. The spectrum was well fit with an absorbed power law spectrum, and it had a hard photon index of 1.26 $\pm$ 0.04, similar to the spectra of known TeV binary systems.

Follow-up observations of HESS J0632+057 at gamma-ray energies above 1 TeV by VERITAS during December 2006 - Jan 2007 and December 2008 - January 2009 yielded upper limits well below the fluxes published by HESS \citep{acciari09b}. The non-detection of HESS J0632+057 by VERITAS together with the HESS results provides evidence for variability in the gamma-ray flux on time scales of months, although the available data do not allow any conclusion on a possible periodicity. 

At optical wavelengths, no reports of variability have been found in the literature. Significant flux variability is reported in radio at 5 GHz on about month time-scales around a mean flux of 0.3 mJy \citep{skilton09}, but no periodic variability could be discerned from these data. This radio flux is orders of magnitude lower than the typical radio flux expected from a TeV blazar which makes that potential interpretation unlikely. 

The point-like nature of the detected TeV source, the excellent positional coincidence with MWC 148 (chance coincidence of $\sim10^{-4}$ according to \citet{aharonian07}), the location on the Galactic plane with a low radio flux, and the variable X-ray and gamma-ray emission from a location coincident with MWC 148 are all facts that argue in favor of an X-ray binary in association with MWC 148. There are currently just three known binary systems that are confirmed as producers of detected TeV gamma ray emission, namely PSR B1259-63 \citep{aharonian05b}, LS 5039 \citep{aharonian05a}, and LS I +61 303 \citep{albert06, acciari08, acciari09}. There is also an accreting black hole binary system, Cygnus X-1, with evidence of detection of a GeV/TeV gamma-ray flare \citep{albert2007}, but this result relies upon marginal significance which would benefit from future confirmation if a high state occurs in the future. The three confirmed TeV binaries are powered by pulsar winds driving shock acceleration or by accretion onto the compact object driving a microquasar jet. Confirmation of HESS J0632+057/MWC 148 as a TeV binary would add a 4th object to this short and special list and allow us to begin studying its interesting properties; whereas, refutation of the binary hypothesis would establish this as an even more mysterious and interesting class of TeV source. The most direct way to test this binary hypothesis is to search for periodic emission signatures. In this paper, we report on recent monitoring data taken with Swift-XRT with the aim of advancing our understanding of this source.

\section{The Observations}
{\it{Swift}} was launched on 2004 November 20 \citep{geh04} with a primary goal to provide detailed measurements of numerous gamma ray bursts (GRBs) and their afterglows with unprecedented reaction times. An ability to slew quickly and frequently has also made {\it{Swift}} an excellent multiwavelength observatory for many non-GRB sources through pre-planned monitoring observations and rapid response target of opportunity (ToO) observations. The narrow field instruments used for these studies are the X-ray telescope (XRT; \citet{burrows05}) and the Ultraviolet-Optical Telescope (UVOT; \citet{roming05}). Bright, hard-spectrum sources can also be studied with the burst alert telescope (BAT) in the $\sim15-150$ keV energy band. For these HESS J0632+057 observations, UVOT was overwhelmed by the bright star in the field of view. The source was not bright enough to be detected by BAT at high energies, or to allow BAT to produce a constraining upper limit. All of the analysis in this paper is based upon XRT data.

The XRT observations were typically taken as 4-6 ksec exposures. The initial set of exposures, from MJD 54857 to MJD 54874, were taken with various sampling intervals chosen primarily to overlap with the simultaneous observations at the VERITAS very high energy gamma ray observatory (see \citet{acciari09b} for results). These observations included a series of 4 observations spaced on daily time intervals, as well as two more observations with 1 week spacing. One month later, starting on MJD 54904, a series of monitoring observations was initiated with $\sim$3-4 day time spacing. This resulted in 20 additional observations extending from MJD 54904 to MJD 54965, at which time the spacecraft had to discontinue observations due to a Sun constraint. The total {\it{Swift}}-XRT data set includes 103 ksec of observations extending from 2009 January 26 to 2009 May 20. All observations beyond January 29 are being reported here for the first time.

\section{Analysis}
The {\it{Swift}}-XRT data were processed using the most recent versions of the standard Swift tools and the most recent calibration files available at the time of data processing. In particular, we utilized Swift Software version 3.0, FTOOLS version 6.6.3, and XSPEC version 12.5.0. Light curves were generated using xrtgrblc version 1.3.

All of the observations were obtained in photon counting (PC) mode. Circular and annular regions are used to describe the source and background areas respectively, and the radii of both regions depend on the measured count rate. In order to handle cases where the sources land on bad CCD detector columns, point spread function correction is handled using xrtlccorr. Since these observations always resulted in XRT count rates $\sim$0.01--0.06 c/s, there was no significant pile-up, which occurs in photon counting mode at count rates above $\sim$0.5 c/s. The full light curves use a bin size of one observation per bin (observations were typically 4-5 ksec duration). We also evaluated single observation light curves, and in these cases, we used a bin size minimum of 20 cts/bin. All error bars are reported at the 1-sigma level, unless otherwise specified.

As described in the results section below, we calculated spectral fits for the entire combined data set. We also calculated spectral fits for a series of observations that were binned temporally with large enough bins to provide enough photon statistics to allow reasonable spectral fits. Following the temporal binning, each observation was fit spectrally using a minimum energy binning ratio of 12 photons/bin, and in most cases the binning ratio was 20 photons/bin. Once the spectral fits were obtained, the flux was calculated. These flux values, which were calculated for each of the temporally binned data segments, were then used to calculate a flux-to-rate ratio during that time period. This ratio was then applied to the rate light curve to convert it to a flux light curve, as shown in Figure \ref{fig:lightcurve_xrt}, which assumes that any spectral variability has a negligible impact on the flux-to-rate ratio within a single time bin.

\section{Results}

\begin{figure}
\includegraphics[scale=0.7]{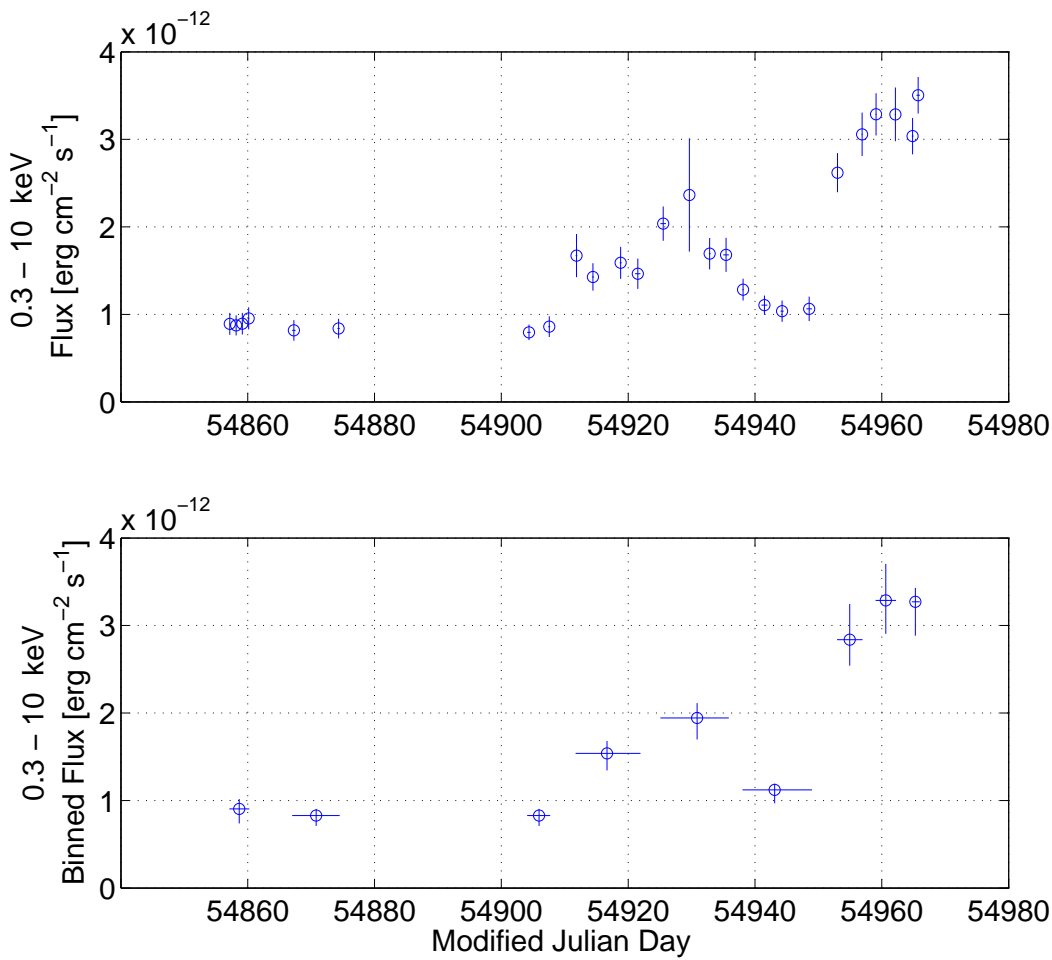}
\caption{The X-ray light curve of XMMU J063259.3+054801 from the {\it{Swift}}-XRT observations in the 0.3-10 keV band. The bottom panel shows the flux derived from binning up all photon events into time bins that are illustrated by the horizontal error bars and finding the best fit spectral model with N$_H$ fixed to 3.1$\times10^{21}$. The flux in the top panel explores temporal variability in more detail by deriving the flux directly from the X-ray rate light curve, with a flux to rate ratio derived for each of the spectral time bins defined by the horizontal error bars in the bottom panel (this assumes that spectral variability is a negligible effect within each time bin).}
\label{fig:lightcurve_xrt}
\end{figure}

The X-ray light curve is shown in Figure \ref{fig:lightcurve_xrt}. There is clear evidence of strong variability on multiple timescales, with measurable flux doubling on timescales as short as $\sim$5 days.  While shorter timescale variability is certainly possible (within the individual $\sim$5 ksec observations), it is not measurable within the flux limited error bars provided by these data. Longer timescale variability is also evident, with an extended rise and decay lasting for at least 30 days within the time frame bracketed by MJD 54907 and MJD 54941. 

\begin{figure}
\includegraphics[scale=0.7]{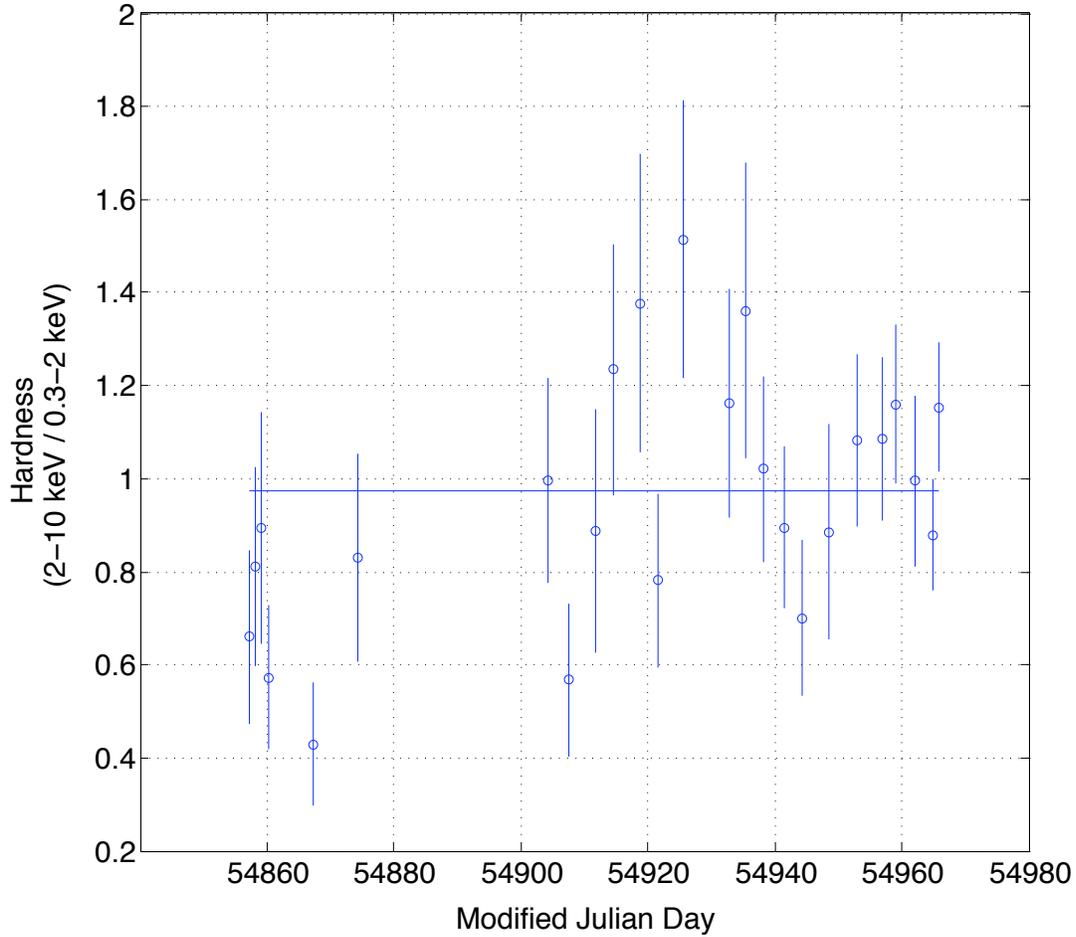}
\caption{The X-ray hardness curve of XMMU J063259.3+054801, where hardness is defined as the flux in the 2-10 keV band divided by flux in the 0.3-2 keV band. The mean hardness is shown as a solid line. A fit to this line produces a reduced $\chi^2$ of 2.01 with 24 degrees of freedom.}
\label{fig:hardness_xrt}
\end{figure}

For this work, the hardness ratio is defined as $R_2/R_1$, where $R_2$ is the rate in the 2-10 keV band and $R_1$ is the rate in the 0.3-2 keV band. The hardness ratio as a function of time is shown in Figure \ref{fig:hardness_xrt}. The mean hardness ratio is 0.96 with a standard deviation of 0.27. Fitting the hardness curve to a constant in time results in a reduced $\chi^2$ equal to 2.01 with 24 degrees of freedom, with a chance probability of 99.8\%. While the constant hardness hypothesis does not provide a good fit, it is not poor enough to claim significant evidence of spectral variability.

\begin{figure}
\includegraphics[scale=0.7]{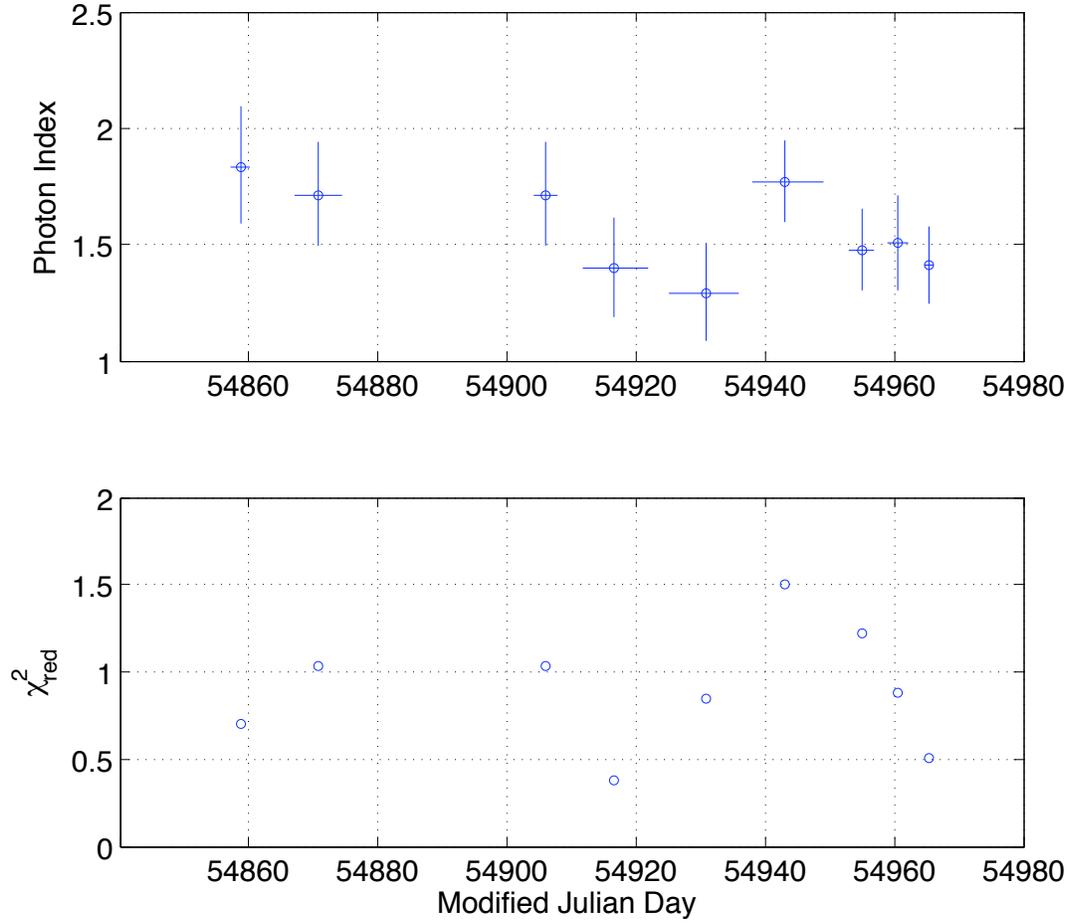}
\caption{Spectral parameters from application of an absorbed power law model to time binned {\it{Swift}}-XRT observations of XMMU J063259.3+054801. The top panel is the power law photon index. The bottom panel is the reduced $\chi^2$ of the model fit. Vertical error bars represent 90\% confidence range.  Horizontal error bars represent the extent of the time region used in the binning, which is the same temporal binning that was shown in Figure 1.}
\label{fig:spectra_NHfixed}
\end{figure}

\begin{figure}
\includegraphics[scale=0.7]{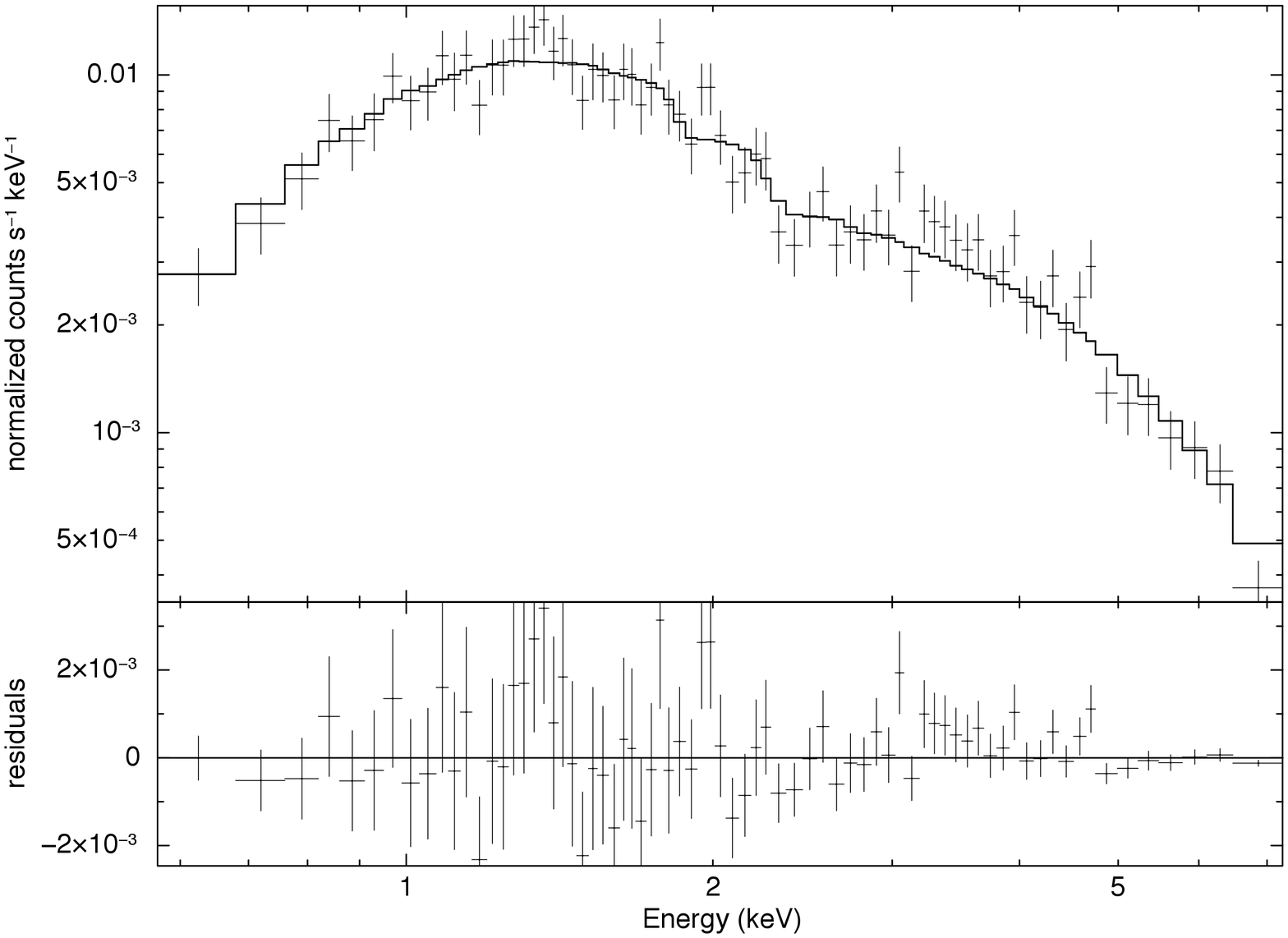}
\caption{Absorbed power law model fit to sum of all {\it{Swift}}-XRT data analyzed in this paper.  $N_H$ is fixed to the value obtained previously by XMM, and the photon index is found to be 1.56$\pm$0.06, with a reduced $\chi^2$ of 0.89. Residuals are shown in the bottom panel.}
\label{fig:overall_spectra_NHfixed}
\end{figure}

A simple absorbed power law model, with $N_H$ left to vary freely, was fit to the time binned data. The $N_H$ was not found to vary significantly. The mean $N_H$ was found to be $3.3\pm0.2 \times10^{21}$ cm$^{-2}$, and all values are consistent, within error, with the value obtained from earlier XMM observations, $3.1\pm0.3 \times10^{21}$ cm$^{-2}$ \citep{hinton09}. For all subsequent fitting, the $N_H$ was fixed to the value obtained from the XMM observations. The photon indices from these fits are shown in Figure \ref{fig:spectra_NHfixed}. The time axis error bars represent the full extent of the time bin used for each model fit, and the vertical error bars are 90\% confidence level error bars. We also binned all of the data together to obtain a combined fit of an absorbed power law to the entire data set. The photon index was found to be 1.56$\pm$0.06, and the fit of the absorbed power law resulted in a reduced $\chi^2$ of 0.89 with 72 degrees of freedom. The model and data for the overall fit are shown in Figure \ref{fig:overall_spectra_NHfixed}.

A power spectrum periodicity search was performed on the overall light curve shown in Figure \ref{fig:lightcurve_xrt}, using a Lomb normalized periodogram which is applicable to unevenly spaced data \citep{press07}. However, no significant periodic signal can be discerned from these data.

\section{Discussion \& Conclusions}

The X-ray flux from the TeV binary LS I +61 303 (which is potentially a very similar object to HESS J0632+057) varies over a 26.5 day orbital cycle at many wavelengths, including variations by factors of 2 to 6 at both optical and X-ray wavelengths \citep{paredes1997, holder07, casares05}, and it has fast timescale flares \citep{smith09, barth08}. Therefore, we have probed HESS J0632+057 for binary orbit variability timescales ranging from days to weeks, as well as searching for any large amplitude flaring on short timescales during all observations. The detection of a binary period would clearly allow us to differentiate between X-ray binary models and Be star wind-driven shock models \citep{townsend2007}, as well as other less likely models such as cosmic rays interacting with a molecular cloud or a remarkably low radio flux blazar on the Galactic plane.

While it is clear that further monitoring observations are needed in order to obtain any conclusions regarding periodicity, it is also clear that the X-ray light curve of this point source is variable on long ($>$ weeks) timescales. The light curve is consistent with periodic variability on timescales longer than those probed in this study, if we assume that there is some additional variability due to flaring, but longer baseline observations are needed to evaluate the reality of any periodic variability. All of the measured fluxes in the epochs during which {\it{Swift}} observed the source are larger (by a factor of $\sim$two) than the flux measured by XMM in September of 2007. A constant flux hypothesis is ruled out. A simplistic sinusoidal curve is also ruled out.  However, it is possible that a more complicated model, involving a long period sinusoidal curve plus flares, could be fit to the light curve. The results of a Lomb-Scargle periodogram analysis show no significant periodicity; however, this analysis does not rule out the possibility of a periodicity at longer timescales beyond the range of these observations ($\ge$54 days) or the possibility of a shorter periodicity timescale with variable or small fractional flux modulation. For example, if the modulation fraction was $\le$30\%, then {\it{Swift}}-XRT would not be sensitive enough to see a significant deviation from one 5 ksec pointing to the next, given the mean flux from HESS J0632+057. It is also likely that this study would miss periodic flux modulation fractions somewhat larger than this, due to the contamination of the periodic signal by non-periodic flare-like emission. Although the following is not statistically significant, we should also mention that if the two high states measured in these data are due to a binary modulation, then the favored period would be approximately 35-40 days; but, of course, there is no evidence for or against the periodic nature of these two high states since longer timescale observations are necessary for making such a determination. If these two high states are not related to any potential periodicity, then the favored period would be longer than half the timeframe of observations performed thus far (i.e. longer than 54 days).

By plotting $N_H/N_{H,gal}$ versus binary orbit for known high mass X-ray binaries, Be-HMXBs and Sg-HMXBs have been found to occupy different regions of parameter space \citep{bodaghee07, rodriguez08}. According to \citet{kalberla2005}, the galactic neutral hydrogen column density, $N_{H,gal}$, is 5.6 $\times$10$^{21}$ cm$^{-2}$ in the direction of HESS J0632+057. If we take the orbital period of HESS J0632+057 to be $\ge$54 days (half the time period of these observations) and the $N_H$ to be $3.1\times10^{21}$ cm$^{-2}$ \citep{hinton09}, then it falls within the middle of the Be-HMXB section of the ÒBodaghee Diagram.Ó This period, as well as the potential $\sim$35-40 day period would both be long enough to be consistent with the interpretation of this object as a TeV emitting HMXB associated with the Be star MWC 148.

If MWC 148 was a single star with a wind driven shock, it would be expected to have a reasonable spectral fit with a thermal model such as an absorbed MeKaL model or an absorbed Raymond-Smith model with a temperature of kT$\sim$1 keV \citep{rauw2002, townsend2007}. However, we find that such models do not fit the data well due to the presence of a harder component. The absorbed Mekal model results in a high temperature of kT=$17\pm4$ keV, which is too high to be likely for a single star model, with reduced $\chi^2$=1.36 for 109 degrees of freedom. The absorbed Raymond-Smith model results in a reduced $\chi^2$=8.15 for 109 degrees of freedom. The spectrum is more naturally fit by an absorbed power law spectrum, which results in a reduced $\chi^2$ of 0.89 and a power law photon index of 1.56$\pm$0.06.  While not definitive, this evidence favors models involving more than a single star, such as the binary hypothesis.

These monitoring observations also provide the possibility to search for short timescale flaring from either XMMU J063259.3+054801 or from any of the other dimmer X-ray sources in the field of view. X-ray flaring has been observed from other X-ray binaries and from at least one TeV emitting X-ray binary, namely LS I +61 303 \citep{smith09}. Short timescale flaring would provide a way to probe the size of the emission region, as well as the power of the engine that must be feeding the associated acceleration site. In principle, short timescale variability could also be used to probe different temporal signatures expected from Be star wind-driven shock models, binary interaction region models, and accretion-driven microquasar jet models. Variability as short as $\sim5$ days was observed, but no flaring at shorter timescales was observed in the flux range probed by these observations.

These data are complemented by the simultaneous data that have been obtained by Fermi, as well as radio observations reported by \citet{skilton09} and TeV observations reported by \citet{acciari09b}.  In future work, modeling of this composite spectral energy distribution and longer timescale observations will provide strong constraints on potential periodicities and on the acceleration site parameters, such as the B field and the electron distribution, implied by fitting a synchrotron spectrum plus an inverse Compton spectrum to the observations.

\acknowledgments
This work is supported at Pennsylvania State University by NASA contract NAS5-00136 and grant NNX08AV77G.




\begin{thebibliography}

\bibitem[Abdo et~al.\ (2009)]{abdo2009} Abdo, A.A. et~al.\ 2009, \apjs, 183, 46
\bibitem[Acciari et~al.\ (2008)]{acciari08} Acciari, V.A. et~al.\ 2008, \apj, 679, 1427
\bibitem[Acciari et~al.\ (2009)]{acciari09} Acciari, V.A. et~al.\ 2009, \apj, 700, 1034
\bibitem[Acciari et~al.\ (2009b)]{acciari09b} Acciari, V.A. et~al.\ 2009b, \apjl, 698, L94 
\bibitem[Aharonian et~al.\ (2008)]{aharonian08} Aharonian, F., Buckley, J., Kifune, T., Sinnis, C. 2008, J.Phys.G: Nucl. Part. Phys., in press
\bibitem[Aharonian et~al.\ (2007)]{aharonian07} Aharonian F., et~al.\ 2007, A\&A, 469, L1
\bibitem[Aharonian et~al.\ (2005a)]{aharonian05a} Aharonian et~al.\ 2005a, Science, 309, 746
\bibitem[Aharonian et~al.\ (2005b)]{aharonian05b} Aharonian et al. 2005b, A\&A, 442, 1
\bibitem[Albert et~al.\ (2007)]{albert2007} Albert, J. et~al.\ 2007, \apjl, 665, L51
\bibitem[Albert et~al.\ (2006)]{albert06} Albert, J. et~al.\ 2006, Science, 312, 1771
\bibitem[Barthelmy et~al.\ (2008)]{barth08} Barthelmy, S. et~al.\ 2008, GCN Circular, 8215
\bibitem[Bodaghee et~al.\ (2007)]{bodaghee07} Bodaghee, A., Courvoisier, T. J., Rodriguez, J., et~al.\ 2007, A\&A, 467, 585
\bibitem[Buckley et~al.\ (2008)]{buckley08} Buckley, J., et~al.\ 2008, APS White Paper on Status of TeV Astronomy; astro-ph/0810.0444
\bibitem[Burrows et~al.\ (2005)]{burrows05} Burrows, D.N., Hill, J.E., Nousek, J.A., et~al.\ 2005, \ssr, 120, 165
\bibitem[Casares et~al.\ (2005)]{casares05} Casares, J. et~al.\ 2005, MNRAS, 360, 1105
\bibitem[Gehrels et~al.\ (2004)]{geh04} Gehrels, N., Chincarini, G., Giommi, P., et~al.\ 2004, \apj, 611, 1005
\bibitem[Hinton et~al.\ (2009)]{hinton09} Hinton, J., Skilton, J., Funk, S., et~al.\ 2009, \apjl, 690, 101
\bibitem[Holder, Falcone, \& Morris (2007)]{holder07} Holder J., Falcone, A., Morris, D. 2007, Proc. 30th Inter. Cosmic Ray Conf., 2, 571
\bibitem[Horan \& Wakely (2009)]{horan09} Horan, D. \& Wakely, S. 2009, Online TeV Catalog; http://tevcat.uchicago.edu
\bibitem[Kalberla et~al.\ (2005)]{kalberla2005} Kalberla et~al.\ 2005, A\&A, 440, 775
\bibitem[Paredes et~al.\ (1997)]{paredes1997} Paredes et~al.\ 1997, A\&A, 320, L25
\bibitem[Press et~al.\ (2007)]{press07} Press, W.H., Teukolsky, S.A., Vetterling, W.T., \& Flannery, B.P. 2007, Numerical Recipes in C, 577 
\bibitem[Rauw et~al.\ (2002)]{rauw2002} Rauw, G., Blomme, R., Waldron, W. L., Corcoran, M. F., Pittard, J. M., et~al.\ 2002, A\&A, 394, 993
\bibitem[Rodriguez \& Bodaghee (2008)]{rodriguez08} Rodriguez, J. \& Bodaghee, A. 2008, Proc. of The Nature and Evolution of X-ray Binaries in Diverse Environments; astro-ph/0801.0961
\bibitem[Roming et~al.\ (2005)]{roming05} Roming, P., et~al.\ 2005, \ssr, 120, 95
\bibitem[Skilton et~al.\ (2009)]{skilton09} Skilton, J., et~al.\ 2009, \mnras, in press; astro-ph/0906.3411
\bibitem[Smith et~al.\ (2009)]{smith09} Smith, A., Kaaret, P., Holder, J., Falcone, A., Maier, G., et~al.\ 2009, \apj, 693, 1621
\bibitem[Townsend et~al.\ (2007)]{townsend2007} Townsend, R. H. D., Owocki, S. P., Ud-Doula, A.\ 2007, \mnras, 382, 139
\bibitem[Weekes (2009)]{weekes09} Weekes, T. 2009, Proc. of 3rd Heidelberg Gamma Ray Symposium

\end{thebibliography}
\end{document}